\documentclass[prb,reprint,twocolumn,amsmath,amsfonts,amssymb]{revtex4}
\usepackage{epsfig}
\def\notes#1{ } 
\def\be{\begin{equation}}
\def\ee{\end{equation}}
\def\bea{\begin{eqnarray}}
\def\eea{\end{eqnarray}}

\def\bra #1 {\langle {#1} \vert}
\def\ket #1 {\vert {#1} \rangle}

\begin{document}
\renewcommand{\thefootnote}{\fnsymbol{footnote}}
\renewcommand{\theequation}{\arabic{section}.\arabic{equation}}

\title{Dual partitioning for effective Hamiltonians to avoid intruders}

\author{Seiichiro Ten-no}
\email[]{E-mail: tenno@garnet.kobe-u.ac.jp}

\affiliation{Graduate School of System Informatics, Department of Computational Sciences, Kobe University, Nada-ku, Kobe 657-8501, Japan}

\date{\today}

\begin{abstract}
We present a new Hamiltonian partitioning which converges an arbitrary number of states of interest in the effective Hamiltonian to the full configuration interaction limits simultaneously.
This feature is quite useful for the recently developed model space quantum Monte Carlo.
A dual partitioning (DP) technique is introduced to avoid the intruder state problem present in the previous eigenvalue independent partitioning of Coope.
The new approach is computationally efficient and applicable to general excited states involving conical intersections.
We present a preliminary application of the method to model systems to investigate the performance.
\end{abstract}

\maketitle
\section{INTRODUCTION}
The quantum Monte Carlo (QMC) methods in configuration space have been increasing the popularity in molecular electronic structure studies in recent years.\cite{afqmc03,afqmc07,ohtsuka08,ohtsuka10,fciqmc09,i-fciqmc,sqmc}
The majority of these approaches are on the basis of the projection,
\be
\psi(\tau)=e^{-\tau(\hat H-E)}\psi(0), \label{eq:proj}
\ee
that converges to the lowest eigen function with a nonzero overlap with the initial wavefunction $\psi(0)$ after a sufficiently long time interval $\tau$.
Several extensions for excited states have been also proposed.\cite{ohtsuka10,fciqmc-excit,MSQMC,mcci,Humeniuk,fciqmc-proj}
One promising approach among them is the model space QMC (MSQMC) based on the effective Hamiltonian formalism.
We shall outline the main features of MSQMC in what follows.

The imaginary-time evolution (ITE) of (\ref{eq:proj}) in infinitesimal time interval can be partitioned into the components, $\psi_{\rm P}(\tau)=\hat P \psi(\tau)$ and $\psi_{\rm Q}(\tau)=\hat Q \psi(\tau)$, as
\be
\frac{d}{d \tau}
\begin{pmatrix}
\psi_{\rm P}(\tau)\\
\psi_{\rm Q}(\tau)\\
\end{pmatrix}
=
-
\begin{pmatrix}
\hat H_{\rm PP} - E & \hat H_{\rm PQ} \\
\hat H_{\rm QP} & \hat H_{\rm QQ} - E
\end{pmatrix}
\begin{pmatrix}
\psi_{\rm P}\\
\psi_{\rm Q}
\end{pmatrix},\label{eq:itmc}
\ee
where $\hat P$ and $\hat Q$ are the projection operators onto the the model space (P-space) and its orthogonal component (Q-space), respectively, and the subscripts of the Hamiltonian mean the individual contributions like $\hat H_{\rm PQ}=\hat P \hat H \hat Q$.
MSQMC treats the model space amplitude $\psi_{\rm P}$ deterministically by the diagonalization of effective Hamiltonian, and the imaginary time evolution of $\psi_{\rm Q}(\tau)$ for a fixed $d\psi_{\rm P}$ is calculated as
\be
\frac { d\psi_{\rm Q}(\tau)}{d\tau} =- (\hat H_{\rm QQ} - E)\psi_{\rm Q}(\tau)-\hat H_{\rm QP}\psi_{\rm P}.\label{eq:itetc}
\ee
We have recently formulated the ITE for the energy dependent partitioning (EDP),\cite{MSQMC}
\be
\frac{d{\bf T}_{\rm QP}(\tau)}{d\tau}=-({\bf H}_{\rm QQ}-{\bf I}_{\rm QQ}E){\bf T}_{\rm QP} -{\bf H}_{\rm QP},\label{eq:itm_tqp_edp}
\ee
as a sufficiency condition of (\ref{eq:itetc}), where ${\bf T}_{\rm QP}$ is the transfer matrix connecting the configuration interaction (CI) coefficients in the P- and Q-spaces as ${\bf C}_{\rm Q}={\bf T}_{\rm QP}{\bf C}_{\rm P}$, and ${\bf I}_{\rm QQ}$ is the identity matrix.
The stationary solution of (\ref{eq:itm_tqp_edp}),
\be
({\bf H}_{\rm QQ}-{\bf I}_{\rm QQ}E){\bf T}_{\rm QP} +{\bf H}_{\rm QP}=0,\label{eq:loewdin}
\ee
corresponds to the L\"owdin partitioning\cite{Loewdin1,Loewdin2} applicable to arbitrary electronic states in the vicinity of $E$ (even with degeneracy).
The dimension of ${\bf T}_{\rm QP}$ can be sizable for a full CI (FCI) problem, and the partitioning technique is usually worthless for determining the exact solutions.
However, the population dynamics of walkers allows us to treat ${\bf T}_{\rm QP}$ stochastically and to integrate the contribution, ${\bf H}_{\rm PQ}{\bf T}_{\rm QA}$, to the effective Hamiltonian on-the-fly.\cite{MSQMC}
Note the ITE (\ref{eq:itm_tqp_edp}) provids a general framework to solve the linear system of equations (\ref{eq:loewdin}) in a stochastic manner.

The MSQMC in EDP is robust and in principle capable of treating arbitrary excited states as has been demonstrated in the calculation of potential energy curves of diatomic molecules, recently.\cite{ohtsuka_ms}
One of the shortcomings of EDP is that independent QMC processes are required for electronic states with different target energies.
Moreover, the computational cost for each of the QMC processes grows linearly with the dimension of the model space.
In the previous paper,\cite{MSQMC} we noted there exists an alternative choice of the eigenvalue independent partitioning (EIP) in MSQMC where all solutions of the effective Hamiltonian converge to the FCI limits simultaneously.\cite{EIP1,EIP2,EIP3}
Nevertheless, EIP seemingly suffers from the intruder state problems as likewise in the usual state-universal multi-reference approaches.
In this communication, we develop a new partitioning to solve all of these points at issue.

\def\np{\noalign{\smallskip}}
\begin{table*}
\begin{center}
\caption
{\label{tab:part}
Properties of the partitionings for effective Hamiltonians.}
\begin{tabular}{lcccccc}
\hline
 && EDP && EIP && DP (present) \\
\hline
Basic partitioning && L\"owdin && Coope && Generalized Coope \\
Exactness && in the vicinity of $E$ && All solutions in ${\bf H}^{\rm eff}_{\rm PP}$ && $M$ solutions in ${\bf H}^{\rm eff}_{\rm PP}$\\
${\bf H}_{\rm PP}^{\rm eff}$ && Symmetric && Nonsymmetric && Nonsymmetric \\
Natural basis for MSQMC && Slater determinants && Eigen functions of ${\bf H}^{\rm eff}_{\rm PP}$ && Eigen functions of ${\bf H}^{\rm eff}_{\rm PP}$\\
Intruder state avoidance && Yes && No && Yes \\
Computational cost && $O(N_{\rm P})$ && $O(N_{\rm P})$ && $O(M)$ \\
\hline
\end{tabular}
\end{center}
\end{table*}

\begin{center}
\begin{figure}[b]
\includegraphics[width=150pt]{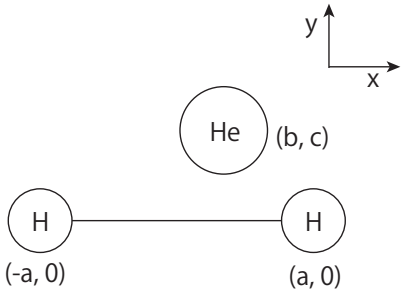}
\caption{The graphical representation of the model system.}
\label{fig:conical}
\end{figure}
\end{center}

\section{THEORY}
Let us consider the standard partitioning of the Schr\"odinger equation,
\be
\begin{pmatrix}
{\bf H}_{\rm PP} & {\bf H}_{\rm PQ}\\
{\bf H}_{\rm QP} & {\bf H}_{\rm QQ}
\end{pmatrix}
\begin{pmatrix}
{\bf C}_{\rm PM} \\
{\bf C}_{\rm QM}
\end{pmatrix}
 = 
\begin{pmatrix}
{\bf C}_{\rm PM} \\
{\bf C}_{\rm QM}
\end{pmatrix}
{\bf \Lambda}_{\rm MM}\label{eq:schb}.
\ee
Here ${\bf \Lambda}_{\rm MM}$ is a diagonal matrix containing state energies.
We assume that the number of solutions of interest $M$ does not exceed the dimension of the P-space, $M \le N_{\rm P}$, and we intend to map (\ref{eq:schb}) onto the effective secular equation in the model space,
\be
{\bf H}_{\rm PP}^{\rm eff} {\bf C}_{\rm PM} = {\bf C}_{\rm PM} {\bf \Lambda}_{\rm MM}\label{eq:schb_eff}.
\ee
For this purpose, we introduce a dual partitioning (DP) further splitting the P-space into the block A with the same dimension as the number of the target states $M$, and the buffer space B as its complement to avoid intruders, ($P=A\cup B$).
The contribution to the effective Hamiltonian from the Q-space is included into the A-component of ${\bf H}_{\rm PP}^{\rm eff}$,
\be
{\bf H}_{\rm PP}^{\rm eff}=
\begin{pmatrix}
{\bf H}_{\rm PA}^{\rm eff}&{\bf H}_{\rm PB}
\end{pmatrix} \label{eq:ab_eff}.
\ee
The comparison of the P-th row of (\ref{eq:schb}), (\ref{eq:schb_eff}), and (\ref{eq:ab_eff}) suggests
\bea
{\bf H}_{\rm PA}^{\rm eff}&=&{\bf H}_{\rm PA}+{\bf \Sigma}_{\rm PA},\\
{\bf \Sigma}_{\rm PA}&=&{\bf H}_{\rm PQ}{\bf T}_{\rm QA}.
\eea
The transfer matrix to relate the CI coefficients as
\be
{\bf C}_{\rm QM}={\bf T}_{\rm QA}{\bf C}_{\rm AM},
\ee
exists as long as ${\bf C}_{\rm AM} $ is invertible.
Formally, the model space equation is contracted into the A-block as
\bea
{\tilde {\bf H}}_{\rm AA}^{\rm eff} {\bf C}_{\rm AM} = {\bf C}_{\rm AM}{\bf \Lambda}_{\rm MM} \label{eq:scha_eff}, \\
{\tilde {\bf H}}_{\rm AA}^{\rm eff}={\bf H}_{\rm AA}^{\rm eff} + {\bf H}_{\rm AB}{\bf t}_{\rm BA},
\eea
with
\be
{\bf t}_{\rm BA}={\bf C}_{\rm BM}{\bf C}_{\rm AM}^{-1}.
\ee
${\bf C}_{\rm QM}$ can be obtained from the linear equation in the Q-th row equation of (\ref{eq:schb}),
\be
{\bf H}_{\rm QQ}{\bf C}_{\rm QM} -{\bf C}_{\rm QM} {\bf \Lambda}_{\rm MM} +{\tilde {\bf H}}_{\rm QM} = 0,\label{eq:cqm}
\ee
where we used an implicit notation for the transformed object indicated by the subscript,
\bea
{\tilde {\bf H}}_{\rm QM}={\tilde {\bf H}}_{\rm QA}{\bf C}_{\rm AM},\\
{\tilde {\bf H}}_{\rm QA}={\bf H}_{\rm QA} + {\bf H}_{\rm QB}{\bf t}_{\rm BA}.
\eea
The elimination of ${\bf \Lambda}_{\rm MM}$ in (\ref{eq:cqm}) using (\ref{eq:scha_eff}) and multiplication of ${\bf C}_{\rm AM}^{-1}$ lead to the corresponding  equation for the transfer matrix,
\be
{\bf H}_{\rm QQ}{\bf T}_{\rm QA} - {\bf T}_{\rm QA}{\tilde {\bf H}}_{\rm AA}^{\rm eff} +{\tilde {\bf H}}_{\rm QA} =0.\label{eq:tqm}
\ee
Either of (\ref{eq:cqm}) and (\ref{eq:tqm}) requires the solution of the model space equation (\ref{eq:schb_eff}), and the equations should be treated by iteration.
Since the linear equation (\ref{eq:cqm}) has exactly the same structure as (\ref{eq:loewdin}) apart from the energy dependency, we can derive the ITE for ${\bf C}_{\rm QM}$ by the analogy of (\ref{eq:itm_tqp_edp}) as
\be
\frac{d{\bf C}_{\rm QM}(\tau)}{d\tau}=-{\bf H}_{\rm QQ}{\bf C}_{\rm QM} +{\bf C}_{\rm QM} {\bf \Lambda}_{\rm MM} -{\tilde {\bf H}}_{\rm QM}.\label{eq:itm_cqm}
\ee
The corresponding equation for the transfer matrix is
\be
\frac{d{\bf T}_{\rm QA}(\tau)}{d\tau}=-{\bf H}_{\rm QQ}{\bf T}_{\rm QA} +{\bf T}_{\rm QA}{\tilde {\bf H}}_{\rm AA}^{\rm eff} -{\tilde {\bf H}}_{\rm QA},\label{eq:itm_tqp}
\ee
which includes couplings between the determinants through ${\bf H}^{\rm eff}_{\rm AA}$, and it is more convenient to use (\ref{eq:itm_cqm}) represented by the $M$ eigen functions of the effective Hamiltonian as the natural basis for an efficient parallelism.
The contribution to the transformed effective Hamiltonian,
\be
{\bf \Sigma}_{\rm PM}={\bf \Sigma}_{\rm PA}{\bf C}_{\rm AM}={\bf H}_{\rm PQ}{\bf C}_{\rm QM},
\ee
can be assembled during the population dynamics for ${\bf C}_{\rm QM}$, and is back transformed by multiplying ${\bf C}_{\rm AM}^{-1}$ to obtain ${\bf H}^{\rm eff}_{\rm PP}$. 

Table \ref{tab:part} summarizes the properties of EDP, EIP, and DP developed in this paper.
For $N_P=1$, all partitionings reduce to EDP.
The present DP becomes identical to EIP of Coope\cite{EIP1} when we are to compute all states as target solutions in the model space, $N_P=M$.
So the present formalism is deemed as a generalization of the Coope partitioning.
It is noted that ${\bf H}^{\rm eff}_{\rm PP}$ in EIP and DP are nonsymmetric while the one in EDP is symmetric.
The nonsymmetric ${\bf H}^{\rm eff}_{\rm PP}$ is the nature of the universality arising from the presence of the overlaps between the real wavefunctions in the Q-space.
On the other hand, the symmetric ${\bf H}^{\rm eff}_{\rm PP}$ in EIP indicates that the eigen functions far from the target energy $E$ are artificial solutions (albeit this is not the necessary condition for EIP).
The intruder state problem occurs from the strong interactions between the states of interest and configuration functions in the Q-space.
And thus EIP suffers from intruders unless all solutions of the effective Hamiltonian are well-separated from the Q-space dominating the real wavefunctions.
The present formalism ameliorate this feature by introducing the flexibility of selecting the roots of interest in the P-space.
Moreover, the number of linear equations to be solved is equivalent to the number of the selected roots $M$ in DP.
This is a significant advancement compared with EDP and EIP, especially when we use a very large model space like complete active space (CAS).

\begin{center}
\begin{figure}[t]
\includegraphics[width=250pt]{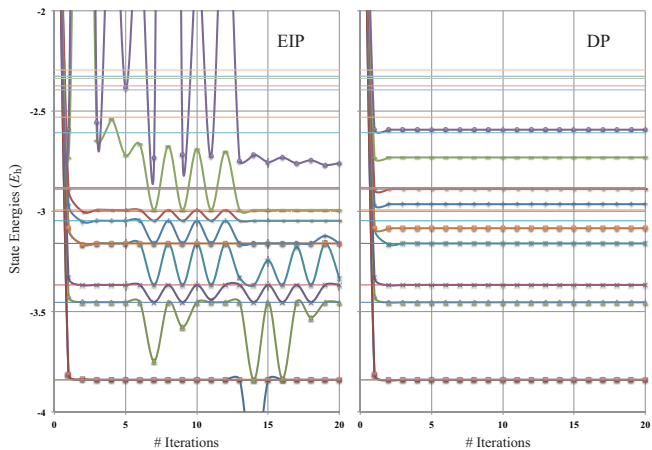}
\caption{The convergence behaviors of state energies for the H$_2$...He model system using EIP and DP. The dimensions of the effective Hamiltonian are $N_{\rm P}=10$ for both EIP and DP, and the number of roots for DP is $M=5$. The FCI limits for the low-lying 20 states are indicated by the thin horizontal lines.}
\label{fig:conv}
\end{figure}
\end{center}

\section{ILLUSTRATIVE EXAMPLES AND DISCUSSIONS}
We test the performance of DP using the model system H$_2...$He as illustrated in FIG. \ref{fig:conical}.
The linear equation is directly solved in a deterministic manner instead of using the stochastic solver of MSQMC, and we investigate the convergence with respect to the number of iterations.
The 6-31G basis set\cite{6-31g} is used, and the geometrical parameters are $a=2$ and $b=c=1$ \AA.
Despite the small FCI dimension (225 in Cs symmetry), the electronic states of the system are rather complex, i.e. there exists a conical intersection between the ground and first excited states, and low-lying excited states exibit near-degeneracies at this geometry.
The model space is constructed by selecting Slater determinants with lower energies in the restricted Hartree-Fock orbital basis.

We show the convergences of EIP and DP in FIG. 2.
The dimension of the model space is $N_{\rm P}=10$, and the EIP result is obtained by setting $M=N_{\rm P}$ of DP.
It is manifested that all 10 solutions spanned by the 10 P-space Slater determinants cannot be not well-separated from the Q-space, and EIP fails to converge.
At the beginning of the iterative cycle in EIP, the oscillatory behaviors are noticeable for higher states, and the transient energies for lower states tend to stay near the FCI values.
As the number of iterations increases, the effect of the intruders is spread in the entire model space, and ${\bf H}^{\rm eff}_{\rm PP}$ collapses.
Contrarily, DP shows a quite stable behavior.
All of the solutions are converged in a few iterations.
By the analogy of the ${\mathcal T}_{1}$ diagnostic in the coupled cluster theory,\cite{T1diag} we also investigate the diagnostic of the transfer matrix,
\be
{\mathcal T}_T = \frac{1}{\sqrt {M}} ||{\bf T}_{\rm QA}||,
\ee
as an indicator of intruders.
${\mathcal T}_T$ are  0.56 and 0.23 for EIP and DP, respectively, after 20 iterations, indicating more significant couplings between P- and Q-spaces for EIP.

To investigate the convergence of DP more precisely, the deviations of the eigen values of ${\bf H}^{\rm eff}_{\rm PP}$ from FCI are listed in TABLE \ref{tab:conv} for the first 7 states.
Among the eigen values in the Table, the first 5 are the target states converging to the FCI energies very rapidly.
The solutions energetically higher than the target states are the buffer ones which converge to the limits different from the FCI energies.
The convergence of the 5th root is slightly slower than the others; this is likely due to the very small gap with the FCI energy of the 6th state.
We examined various systems to obtain essentially similar behaviors.

The reconstruction of the model space has not been considered so far.
The convergence of DP can be deteriorated when a target sate is strongly coupled with a Slater determinant in the Q-space.
In this case, DP can easily include that determinant into the P-space as in the promotion step of MSQMC.\cite{MSQMC}
The promotion of EIP can introduce additional intruders.
For the stochastic solver based on (\ref{eq:itm_cqm}), the only modification of the MSQMC algorithm from EDP is in the P-spawning step, in which a progeny is generated from a linear combination of Slater-determinants rather than a single determinant.
In addition, DP might be also useful to formulate multi-reference theories to avoid intruders.

\section{CONCLUSIONS}
In this note, we developed the DP technique for effective Hamiltonians.
DP is considered to be the generalization of EIP of Coope, and is capable of providing $M$ roots simultaneously avoiding intruders.
DP requires to solve $M$ linear system of equations.
This is a significant advantage over EDP of L\"owdin, which requires $N_{\rm P}$ linear equations for each root, e.g. DP is 10 times as fast as EDP even for a small problem, $M=2$ and $N_{\rm P}=10$.
We therefore propose to employ DP as the standard partitioning for MSQMC.
Recently, Overy et al. proposed a way to calculate pure expectation values introducing a replica population of walker dynamics in FCIQMC.\cite{replica}
It is also straightforward to extend the replica technique to MSQMC with DP to calculate density matrices for various transition properties like transition moments and circular dichroism spectra.
We will report on the works along this line in the near future.

\begin{widetext}
\begin{center}
\begin{table}
\caption
{\label{tab:conv}
Convergence of state energies ($E_{\rm h}$) in the present ${\bf H}^{\rm eff}_{\rm PP}$ for $M=5$ and $N_{\rm P}=10$. Values less than $10^{-6}$ are denoted by dots.}
\begin{tabular}{ccrrrrrrrrrrrrrrrrrr}
\hline
&&& \multicolumn{9}{c}{Target solutions} & \multicolumn{7}{c}{Buffer solutions} \\
\cline{4-12}\cline{14-19}
&Iteration && 1 && 2 && 3 && 4 && 5 && 6 && 7 && -- & \\
\hline\hline
& 1 && -0.024968 && -0.026310 && -0.031164 && -0.036053 && -0.075830 && -0.075922 && -0.082326 && -- \\
& 2 && 0.000342 && 0.000004 && -0.000762 && 0.000572 && 0.004270 && -0.075814 && -0.082568 && -- \\
& 3 && -0.000005 && -0.000016 && -0.000124 && -0.000019 && -0.000285 && -0.075810 && -0.082652 && -- \\
& 4 && $\cdots$ && -0.000001 && -0.000015 && -0.000001 && -0.000006 && -0.075810 && -0.082686 && -- \\
& 5 && $\cdots$ && $\cdots$ && $\cdots$ && $\cdots$ && -0.000008 && -0.075811 && -0.082699 && -- \\
& 6 && $\cdots$ && $\cdots$ && $\cdots$ && $\cdots$ && -0.000003 && -0.075811 && -0.082705 && -- \\
& 7 && $\cdots$ && $\cdots$ && $\cdots$ && $\cdots$ && -0.000001 && -0.075811 && -0.082707 && -- \\
& 8 && $\cdots$ && $\cdots$ && $\cdots$ && $\cdots$ && $\cdots$ && -0.075811 && -0.082708 && -- \\
& 9 && $\cdots$ && $\cdots$ && $\cdots$ && $\cdots$ && $\cdots$ && -0.075811 && -0.082708 && -- \\
\hline
& FCI && -3.840350 && -3.840340 && -3.453751 && -3.366866 && -3.160320 && -3.160116 && -3.047064 && -- \\
\hline
\end{tabular}
\end{table}
\end{center}
\end{widetext}

\begin{acknowledgements}
This work was partly supported by the Strategic Programs for Innovative Research (SPIRE) and FLAGSHIP2020 as the priority issue 5 (Development of new fundamental technologies for high-efficiency energy creation, conversion/storage and use) of MEXT, and the Computational Materials Science Initiative (CMSI). We are also grateful for the computer resources through the HPCI System Research project (Project ID: hp150228, hp150278). 
\end{acknowledgements}


\begin{thebibliography}{99}
\bibitem{afqmc03} S. Zhang, H. Krakauer, {\it Phys. Rev. Lett.}, \textbf{90}, 136401 (2003).
\bibitem{afqmc07} W. A. Al-Saidi, S. Zhang, H. Krakauer, {\it J. Chem. Phys.}, \textbf{127}, 144101 (2007).
\bibitem{ohtsuka08} Y. Ohtsuka, S. Nagase, {\it Chem. Phys. Lett.}, \textbf{463}, 431 (2008).
\bibitem{ohtsuka10} Y. Ohtsuka, S. Nagase, {\it Chem. Phys. Lett.}, \textbf{485}, 367 (2010).
\bibitem{fciqmc09} G. H. Booth, A. J. W. Thom, A. Alavi, {\it J. Chem. Phys.}, \textbf{131}, 054106 (2009).
\bibitem{i-fciqmc} D. Cleland, G. H. Booth, A. Alavi, {\it J. Chem. Phys.}, \textbf{132}, 174104 (2010).
\bibitem{sqmc} F. R. Petruzielo, A. A. Holmes, H. J. Changlani, M. P. Nightingale, C. J. Umrigar, {\it Phys. Rev. Lett.}, \textbf{109}, 230201 (2012).
\bibitem{fciqmc-excit} G. H. Booth, G. K.-L. Chan, {\it J. Chem. Phys.}, \textbf{137}, 191102 (2012).
\bibitem{MSQMC} S. Ten-no, {\it J. Chem. Phys.}, \textbf{138}, 164126 (2013).
\bibitem{mcci} J. P. Coe, M. J. Paterson, {\it J. Chem. Phys.}, \textbf{139}, 154103 (2013).
\bibitem{Humeniuk} A. Humeniuk, R. Mitri\'c, {\it J. Chem. Phys.}, \textbf{141}, 194104 (2014).
\bibitem{fciqmc-proj} N. S. Blunt, S. D. Smart, G. H. Booth, A. Alavi, {\it J. Chem. Phys.}, \textbf{143}, 134117 (2015).
\bibitem{Loewdin1} P.-O. L\"owdin, {\it J. Chem. Phys.}, \textbf{19}, 1396 (1951).
\bibitem{Loewdin2} P.-O. L\"owdin, {\it J. Mol. Spec.}, \textbf{14}, 112 (1964).
\bibitem{ohtsuka_ms} Y. Ohtsuka, S. Ten-no, {\it J. Chem. Phys.}, \textbf{143}, 214107 (2015).
\bibitem{EIP1} J. A. R. Coope, {\it Mol. Phys.} \textbf{18}, 571 (1970).
\bibitem{EIP2} J. A. R. Coope, D. W. Sabo, {\it J. Comput. Phys.} \textbf{23}, 404 (1977).
\bibitem{EIP3} D. Mukhopadhyay, S. Mukhopadhyay, R. Chaudhuri, D. Mukherjee, {\it Theor. Chim. Acta} \textbf{80}, 441 (1991), and references therein.
\bibitem{6-31g} W. J. Hehre, R. Ditchfield, J. A. Pople, {\it J. Chem. Phys.} \textbf{56}, 2257 (1972).
\bibitem{T1diag} T. J. Lee, J. E. Rice, G. E. Scuseria, H. F. Schaefer, {\it Theor. Chim. Acta.}, \textbf{75}, 81 (1989).
\bibitem{replica} C. Overy, G. H. Booth, N. S. Blunt, J. J. Shepherd, D. Cleland, A. Alavi, Ali, {\it J. Chem. Phys.}, \textbf{141}, 244117 (2014).
\end{thebibliography}
\end{document}